\documentstyle[preprint,aps]{revtex}       
\begin{document}
\preprint{\vbox{Published in Physica Scripta 59, 352 (1999)}}
\title{Centroids of Gamow-Teller transitions at finite temperature in
fp-shell neutron-rich nuclei}
\author{O. Civitarese $^{1}$\footnote{Email : 
civitare@venus.fisica.unlp.edu.ar},
and A. Ray $^{2}$\footnote{Email : akr@tifr.res.in } 
\footnote{Current address: Tata Institute of Fundamental
 Research, Bombay 400 005, India}}
\address{$^1$ Department of Physics, University of La Plata,
C.C.67, La Plata (1900), Argentina \\
$^2$ Laboratory for High Energy Astrophysics, Code 661, \\
    NASA/Goddard Space Flight Center, Greenbelt, MD 20771, USA }
\date{\today}
\maketitle
\begin{abstract}

The temperature dependence of the energy centroids and strength 
distributions for
Gamow-Teller (GT) $1^+$ excitations in several fp-shell nuclei is studied. 
The quasiparticle
random phase approximations (QRPA) is extended to describe GT states at 
finite temperature.
A shift to lower energies of the GT$^+$ strength is found, as compared to 
values obtained at zero temperature. 

\end{abstract}

Weak-interaction mediated reactions on nuclei in the core of massive stars
play an important role in the evolutionary stages leading to a
 type II supernova. These
reactions are also involved in r-process nucleosynthesis \cite{a}. Nuclei in
the fp-shell participate in these reactions in the post-
silicon burning stage of a pre-supernova star \cite{b}. The astrophysical
scenarios, where these reactions can take place,
depend upon various nuclear structure related quantities \cite{c}.
Among them, the energy  centroids for GT and IAS  transitions can 
determine the yield of electron and neutrino captures.
The dependence of such nuclear observables upon 
the stellar temperature is a matter of interest \cite{d}. In the present
letter we are addressing the question about the temperature dependence of the
energy-centroids of GT$^{\pm}$ transitions \cite{e}. We have performed 
microscopic calculations of these centroids
using the finite-temperature quasiparticle random phase approximation
\cite{f} and for 
temperatures (T) below critical values related with the collapse of 
pairing gaps (T$\leq $1 MeV )\cite{g}. These temperatures are 
near the values characteristic of the pre-supernova core \cite{d}.
The consistency of the approach has been tested by 
evaluating, at each temperature, the Ikeda Sum Rule and total
GT$^-$ and GT$^+$ strengths \cite{h}. \par 

The starting Hamiltonian is 

\begin{equation}
H=H_{\rm{sp}}+H_{\rm{pairing}}+H_{\rm{GT}} \,
\end{equation}
 
where by the indexes (sp),(pairing) and (GT) we are denoting the single-particle,
pairing and Gamow-Teller ($\sigma \tau$.$\sigma \tau$) terms, respectively. 
For the pairing interaction, both for protons and neutrons, a separable monopole
force is adopted with coupling constants $G_p$ and $G_n$ and for the residual
proton-neutron Gamow-Teller interaction $H_{\rm{GT}}$ the form given by
Kuzmin and Soloviev \cite{i} is  assumed. As shown in the context of nuclear
double $\beta$ decay studies \cite{j} the Hamiltonian (1) reproduces
the main features found in calculations performed with realistic interactions.
The structure of the residual interaction-term can be defined as the sum of
particle-hole ($\beta^{\pm}$) and particle-particle ($P^{\pm}$) 
terms of the $\sigma {\tau}^{\pm}$ operators, as shown in \cite{i}
\cite {j}, namely:

\begin{equation}
H_{\rm{GT}}=2\chi (\beta^-\beta^+)-2 \kappa(P^-P^+)    \,
\end{equation}
in standard notation.

To generate the spectrum of $1^+$ states associated with the Hamiltonian (1) we have
transformed it to the quasiparticle basis, by performing BCS transformations
for proton and neutrons channels separately, and then diagonalized the
residual interaction between pairs of quasiprotons (p) and quasineutrons (n) in the
framework of the pn-QRPA \cite{j} \cite{l}. This procedure leads to the definition
of phonon states in terms of which one can write both the wave functions and the
transition matrix elements for $\sigma \tau^-$ and $\sigma \tau^+$ excitations
of the mother nucleus. Since the procedure can be found in textbooks 
we shall omit further
details about it and rather proceed to the description of the changes which are needed
to account for finite temperature effects. Like before one  has to treat pairing
correlations first, to define the quasiparticle states at finite temperature, and
afterwards transform the residual interaction into this basis to diagonalize
the pn-QRPA matrix. The inclusion of thermal effects on the pairing Hamiltonian
is performed by considering thermal averages in dealing with the BCS equations.
Details of this procedure can be found in \cite{g}. The most notable effect of thermal
excitations on the pairing correlations is the collapse of the pairing gaps, at temperatures
of the order of half the value of the gap at zero temperature. For a separable pairing force
the finite temperature gap equation is written \cite{g}

\begin{equation}
\Delta(T)=G\sum_{\nu}u_{\nu}v_{\nu}(1-2f_{\nu}(T))   \,
\end{equation}

where the factors $f_{\nu}(T)= {(1+e^{E_{\nu}/T})}^{-1}$ are the thermal occupation factors for
single quasiparticle states. The quasiparticles energies $E_{\nu}=\sqrt{{(e_{\nu}-\lambda)}^2+
{\Delta(T)}^2}$ are now functions of  the temperature, as well as the occupation factors
$u$ and $v$.

The thermal average procedure of \cite{g} accounts for the inclusion of excited states
in taking expectation values at finite temperature. It has also been used to describe
two-quasiparticle excitations and QRPA states at finite temperature \cite{f}. In the
basis of unlike(proton-neutron)-two-quasiparticle states the thermal average gives,
for the commutator between pairs, the expression:

\begin{equation}
 < [ {\alpha}_{\nu,n} {\alpha}_{\mu,p} , {\alpha}_{\rho,p}^{\dagger}
 { \alpha}_{\gamma,n}^{\dagger} ] >=
 {\delta}_{\nu,\gamma}{\delta}_{\mu,\rho}(1-f_{\nu,n}-f_{\nu,p}) 
\end{equation}

These factor have to be included in the pn-QRPA equations \cite{l} in taking the
commutators
which lead to the pn-QRPA matrix, as they have been considered in dealing with pairs of
like-(neutrons or protons)-quasiparticles \cite{f}. More details about this procedure, for the
particular case of proton-neutron excitations, will be given in \cite{p}.

The single particle basis adopted for the present calculations consists of
the complete f-p and s-d-g shells and the
corresponding intruder state $0h_{11/2}$, both for protons and neutrons.
In this single particle basis, with energies obtained from a fit of the
observed one-particle spectra at the begining of the fp-shell, and taking
$^{40}$Ca as an inert core we have solved temperature dependent BCS equations
\cite{g} for temperatures 0~MeV $\leq$ T $\leq$ 0.8~MeV. The coupling constants
$G_p$ and $G_n$, of the proton and neutron monopole pairing channels of 
Eq.(1), have been fixed to reproduce the experimental data on
odd-even mass differences. In Table 1 the 
calculated neutron and proton pairing gaps at T=0~MeV are compared to
the experimental values extracted from  \cite{k}. Once the pairing coupling 
constants are determined one can calculate the standard zero temperature 
pn-QRPA \cite{j} \cite{l} equations of motion to reproduce the known systematics \cite{e}
of GT$^{\pm}$ energies and strengths.
>From the comparison between the calculated and
experimental values for GT$^{\pm}$ energies and B(GT$^{\pm}$) strengths we have 
fixed the values of the coupling constants $\chi$ and $\kappa$ of the
Hamiltonian Eq.(2). \par

The consistency of the
pn-QRPA basis is also given by the ratios between the calculated and
expected values of the Ikeda's sum rule 3(N-Z). The values of the 
above quantities are shown in Table 2. 
The experimental values of the B(GT$^+$) strengths have been approximated by using the
expression \cite{m} 

\begin{equation}
\frac{\rm{B(GT}^+)} {\rm{Z}_{\rm{valence}}(20-\rm{N}_{\rm{valence}})}=a +
b (20-\rm{Z}_{\rm{valence}})\rm{N}_{valence}   \,
\end{equation}

where a=3.48 10$^{-2}$ and b=1.0 $10^{-4}$ (see also \cite{n}).

The overall agreement between calculated and 
experimentally determined values at zero temperature, both for pairing and GT observables, is
rather good. We are now in a position to calculate these observables at finite
temperature. At a given value of T we have solved the pairing gap equations and the
pn-QRPA equations. With the resulting spectrum of $1^+$ states, both
for GT$^-$ and GT$^+$ excitations, and the corresponding transition matrix elements 
of the $\sigma \tau^+$ operator we have obtained the values shown in Table 3, where
from the energy-centroids

\begin{equation}
\rm{E(T)}=\frac{\sum_n \rm{E}_n(1^+) {\mid <1^+_n \mid \mid \sigma
 \tau^+ \mid \mid g.s >
\mid}^2}
{\sum_n{\mid <1^+_n \mid \mid \sigma \tau^+ \mid \mid g.s >
\mid}^2}     \,
\end{equation}

we have extracted the temperature dependent shifts

\begin{equation}
\delta_{\rm{E}}(\rm{T})= \rm{E(T=0)}-\rm{E(T)}
\end{equation}

Since the changes of the calculated 
centroids for GT$^-$ excitations at different temperatures are minor
we are showing only the quantities which correspond to GT$^+$ transitions.
Let us discuss some features shown by the result of the present calculations
by taking the case of $^{56}$Fe as an example.
As known from previous studies \cite{l}, the repulsive effects due to 
the proton-neutron residual interactions affect
both the GT$^-$ and the GT$^+$ unperturbed strength distributions,
moving them up to higher energies. The large upwards-
shift, as compared to the strength distribution of the unperturbed
proton-neutron two-quasiparticle states, is exhibited by the GT$^-$
distribution \cite{l}. At finite temperatures two different effects become important,
namely: the vanishing  of the pairing gaps and the thermal blocking 
of the residual interactions. In order to distinguish between both effects we
have computed GT-strength distributions for the case of the unperturbed
proton-neutron two-quasiparticle basis. The pairing gaps, for
proton and neutrons, collapse at temperatures T$\approx$ 0.80 MeV.
At temperatures below these critical values (T= 0.7 MeV)
the neutron and pairing gaps decrease to about $50\%$ and $40\%$ of the 
corresponding values at T=0, respectively. At this temperature (T=0.7 MeV) these changes
amount to a lowering of the centroid for GT$^+$ transitions  of the order 
of 1 MeV. When the residual interaction is turned on the resulting shift
to lower energies is $\approx $1.20~MeV.      
>From these results it can be seen that the total downward shift of the GT$^+$
centroid is not solely due to pairing effects but also due to the thermal
blocking of the proton-neutron residual interactions.
This result can be understood as follows. Since GT$^+$
transitions are naturally hindered by the so-called Pauli blocking effect
the smearing out of the Fermi surface due to pairing correlations, at zero temperature, 
tends to favour them. When temperature
dependent averages are considered, Eq.(3), the pairing correlations are gradually
washed-out as the temperature increases. This in turn leads to a 
sharpening of the Fermi surface thus decreasing the value of the energy of the unperturbed 
proton-neutron pairs. In addition, from the structure of
the pn-QRPA equations at finite temperature, it can easily be seen that 
factors such as in Eq.(4) will appear screening the interaction terms.
This additional softening of the repulsive GT interaction
adds to the decrease of the unperturbed proton-neutron energies and the result is a
larger shift of the GT$^+$ centroids. It should be noted that the position of the 
centroid of the GT$^-$ transitions is less sensitive to these effects,
as we have mentioned before. The calculated shifts for these centroids are 
of the order of (or smaller than) 0.5 MeV.

This result, concerning GT$^-$ centroids, is understood by noting that
the collapse of the proton pairing gap does not affect the BCS unoccupancy
factor ($u_p$) for proton levels above the Fermi surface as well as the
BCS occupancy factor ($v_n$) for neutron levels below the Fermi surface and the
energy of the unperturbed proton-neutron pairs remains nearly the same.   
Table 3 shows similar features for the changes of the GT$^+$ energy-centroids 
in other cases. \par

To conclude, in this work we have presented the result of temperature dependent QRPA
calculations of GT transitions in some neutron-rich nuclei in the fp-shell.
The energy centroids of these transitions have been calculated at temperatures
below the critical values associated with the collapse of the pairing
correlations. The inclusion of thermal averages on the QRPA 
equations of motion leads to the softening of the repulsion induced by the 
Gamow-Teller interaction
on proton-neutron pairs as well as to the sharpening of the proton and
neutron Fermi surfaces. The combined effects of both mechanisms leads to
a downwards-shift of the GT$^+$ strength while the GT$^-$ centroids remain
largely unaffected.  
We have observed the constancy of the total GT$^+$ strength, as a 
function of
temperature, in agreement with previously reported
results of the Monte Carlo Shell Model Method \cite{o}.   
The shift of the GT$^+$ centroids at finite temperatures will effectively 
lower the energy thresholds for 
electron capture reactions in stellar enviroments leading to more
neutronization at lower stellar densities during gravitational collapse. On the other hand 
the neutrino
induced r-process reactions will remain relatively unaffected by the 
small ( $\leq $ 0.5 MeV)  thermally induced shifts of GT$^-$ centroids.
Considering that the empirically determined
energies of the GT$^+$ centroids are known with an accuracy of the order of 0.43 MeV \cite{e}
the temperature-dependent effects reported here may be significant for 
astrophysical rate calculations and their consequences for pre-supernova stellar
evolution and
gravitational collapse.
Work is in progress to predict GT$^+$(GT$^-$) centroids and
strengths, by using the pn-QRPA method at finite T, for nuclei for 
which experimental data 
through charge-exchange (p,n)( (n,p))-reactions
in the opposite direction is available to constrain the 
former one. \cite{p} \par

We thank George Fuller for fruitful discussions and
the Institute for Nuclear Theory at the University
of Washington for its hospitality and the DOE for partial
support during the completion of this work. (O.C) is a fellow of the CONICET
(Argentina) and acknowledges the grant ANPCYT-PICT0079; (A. R) is a (U.S.) 
National Research Council sponsored Resident Research 
Associate at NASA/Goddard Space Flight Center.\par

\eject

\eject

\begin{table}[t]
\begin{center}

\caption{Experimental and calculated pairing gap parameters (in MeV) 
at zero temperature. The experimental values, extracted from ref.$^{11)}$, 
are listed in parenthesis.}

\begin{tabular}{ccc}
\hline
\hline
 Nucleus    & Neutron Gap  & Proton Gap   \\
 $^{54}$Fe & 1.647~(1.694) & 1.560~(1.520)  \\
 $^{56}$Fe & 1.359~(1.363) & 1.593~(1.568) \\
 $^{58}$Ni & 1.262~(1.298) & 1.752~(1.725) \\
 $^{60}$Ni & 1.469~(1.489) & 1.743~(1.726) \\
\hline
\end{tabular}
\end{center}
\end{table}

\eject

\begin{table}[t]
\begin{center}

\caption{Experimental and calculated energy centroids and total
strenghts for GT($^{\pm}$) transitions, at zero  temperature.
The available experimental values are listed in parenthesis. 
The energies (E) correspond 
to excitations from the ground state of the mother nucleus. 
The experimental values of the B(GT$^-$)-strengths (third column, in
parenthesis) are taken from ref. $^{17)}$ and the experimental 
values of the B(GT$^+$)-strenghts (fourth column, in parenthesis) have been 
obtained by using Eq.(5) (ref.$^{13)}$).
The last column shows the ratio between calculated (B(GT$^-$)-B(GT$^+$)) and
expected (3(N-Z)) values of the Ikeda Sum Rule.}

\begin{tabular}{ccccc}
\hline
\hline
  Nucleus  &  E(GT$^-$) (MeV) & B(GT$^-$) &  B(GT$^+$)     &  Ratio  \\
 $^{54}$Fe &  8.91~(8.90) &  9.290~(7.8 $\pm$ 1.9) &  3.306~(3.312)  &  0.997  \\
 $^{56}$Fe &  9.31~(9.00) & 14.630~(9.9 $\pm$ 2.4)  &  2.632~(2.928)  &  0.999 \\
 $^{58}$Ni &  9.48~(9.40) & 10.860~(7.4 $\pm$ 1.8)  &  4.836~(3.744)  &  1.000 \\
 $^{60}$Ni &  9.22~(9.00) & 14.890~(7.2 $\pm$ 1.8) &  2.890~(3.148)  &  1.000  \\
\hline

\end{tabular}

\end{center}

\end{table}

\eject

\begin{table}[t]
\begin{center}

\caption{Calculated values of the shift $\delta_{\rm{E}}(\rm{T})$,  
Eq.(7), of the energy centroid for GT$^+$ excitations for different values of
the temperature (T). 
The experimentally determined energy centroids, (E$_{\rm{exp}}$), 
are taken from the compilation
of data given in ref.$^{5)}$. All values are given in units of MeV.} 
\begin{tabular}{ccccc}
\hline
\hline
Nucleus & E$_{\rm{exp}}$ (MeV)  & T=0.4 MeV & T =0.6 MeV & T=0.8 MeV  \\
$^{54}$Fe & 3.5       & 0.179    & 0.745    & 1.483     \\
$^{56}$Fe & 6.0       & 0.216    & 0.807    & 2.038     \\
$^{58}$Ni & 3.7       & 0.071     & 0.342     &0.955       \\
$^{60}$Ni & 5.4       & 0.118    & 0.450     & 0.932      \\
\hline
\end{tabular}
\end{center}
\end{table}

\eject

\end{document}